# Analysis of Carrier Accumulation in Active Region by Energy Loss Mechanisms in InGaN Light-Emitting Diodes

Dong-Pyo Han[1, 2, a)], Jong-In Shim[2], and Dong-Soo Shin[3]

[1]*Faculty of Science and Technology, Meijo University, 1-501 Shiogamaguchi, Tempaku-ku, Nagoya 468-8502, Japan*

[2]*Dept. of Electronics and Communication Engineering, Hanyang University, ERICA Campus, Ansan, Gyeonggi-do 426-791, Korea*

[3]*Dept. of Applied Physics and Dept. of Bionanotechnology, Hanyang University, ERICA Campus, Ansan, Gyeonggi-do 426-791, Korea*

**Abstract**

Carrier recombination and transport processes play key roles in determining the optoelectronic performances such as the efficiency droop and forward voltage in InGaN/GaN multiple-quantum-well (MQW) light-emitting diodes (LEDs). In this work, we investigate the dominant carrier transport and recombination processes inside and outside the MQW region as a function of injection current from a new point of view by separately examining the carrier energy loss. Analysis of the measurement results reveals that the carrier accumulation and subsequent spill-over from the MQW active region to the clad is the most probable mechanism of explaining the efficiency and forward voltage variation with the injection current.

a) E-mail: han@meijo-u.ac.jp

In a recent decade, significant progress has been made in the development of high-efficiency InGaN-based light-emitting diodes (LEDs) covering from the visible to the near-ultraviolet spectral ranges.[1,2] The InGaN-based LEDs, however, still suffers from a reduction (droop) of the internal quantum efficiency (IQE) and high operating voltage with increasing driving current while many applications require high wall-plug efficiency (WPE).[3,4] The efficiency droop and high operating voltage phenomenon is currently a key subject of intense research as its solution can expedite the general lighting with white-light sources based on the InGaN LEDs.

Many physical mechanisms have been reported as possible origins of the efficiency droop in InGaN LEDs, including the saturation of the radiative recombination rate and the subsequent increase of the nonradiative recombination rate via carrier spill-over from multiple quantum wells (MQWs) to the p-GaN clad,[5,6] direct hot electron transport to p-GaN,[7,8] asymmetric carrier distribution in MQWs,[9,10] and carrier losses inside and/or outside the MQWs by the Auger recombination process.[11,12] The limited radiative recombination rate in the InGaN active layer is explained as possibly caused by a small electronic density of states (DOS) due to piezo electric filed,[4] local potential fluctuations in In-containing quantum wells (QWs),[5,6] insufficient carrier capture rate,[7,8] and high energy excitation via Auger processes.[11,12] Most of the proposed models consider that the nonradiative recombination outside the MQW active region is mainly responsible for the efficiency droop in InGaN-based LEDs. However, a systematic methodology confirming the models has not yet been known.

In this paper, we investigate the carrier recombination and transport processes in InGaN/GaN MQW LEDs by examining the carrier energy loss mechanisms as a function of driving current. Several carrier energy loss mechanisms are considered including the ohmic loss, photon absorption, the Shockley-Read-Hall (SRH) recombination, and the Thomson heating and cooling effects. Through the analysis of experimental results, it is revealed that the carrier accumulation and subsequent spill-over and/or leakage from the MQWs is the possible origin of efficiency droop and high operating voltage, which process begins from a relatively low current density.

For experiments, we used an InGaN/GaN MQW-based blue LED (emission wavelength: ~450 nm at 300 K) grown on a *c*-plane sapphire substrate by the metal-organic chemical vapor deposition (MOCVD). A conventional epitaxial structure was utilized, consisting of five pairs of MQW active layer and an AlGaN electron-blocking layer (EBL). The chip was fabricated with lateral-type electrodes (chip size: 900×600 μm$^2$) and mounted on a surface-mount-device (SMD) package. A Keithley semiconductor parameter analyzer under the pulsed-voltage condition was used for electrical analysis. The absolute light output power was measured by an integrating sphere under the pulsed-current driving condition (pulse period: 100 μs, duty cycle: 1%) to minimize the self-heating effect. Both the IQE and the light extraction efficiency were obtained by using the temperature-dependent electroluminescence (TDEL) measurement, which is widely used in various research works.[5,13,14]

Figure 1 depicts the electrical input power ($P_{elec}$), the absolute light output power ($P_{opt}$), the electrical power loss ($P_{loss}$), and the wall-plug efficiency ($\eta_{WPE}$) as a function of total driving current (*I*) of an LED under investigation. $P_{elec}$ is the total electrical input power which is a product of driving current *I* and applied voltage *V* and $\eta_{WPE}$ is defined as $\eta_{WPE} = P_{opt}/P_{elec}$. $P_{loss}$ is obtained experimentally from the $\eta_{WPE}$ data as shown below:

$$P_{loss} = (1 - \eta_{WPE})P_{elec}. \qquad (1)$$

$P_{loss}$ represents the electrical power lost in the LED by various processes discussed below in more detail. In Fig. 1, $P_{loss}$ increases rapidly and becomes larger than $P_{opt}$, which is the so-called efficiency droop.

$\eta_{WPE}$ is composed of four efficiencies such as the voltage efficiency ($\eta_{VE}$), the radiative efficiency ($\eta_{RE}$), the injection efficiency ($\eta_{IE}$), and the light-extraction efficiency ($\eta_{LEE}$).[15] Since each efficiency has a different energy loss mechanism, it is very informative to analyze $P_{loss}$ according to each energy loss mechanism. Four power loss mechanisms are considered here, namely, the Joule heating during the carrier transport ($P_{Joule}$), the total electrical power loss through the nonradiative carrier recombination processes ($P_{NR}$), photon absorption caused by nonescaping from the inside of an LED chip ($P_{abs}$), and carrier heating and cooling between the wells and barriers ($P_{HC}$).[16] $P_{Joule}$, $P_{NR}$, $P_{HC}$, and $P_{abs}$ are dominant loss mechanisms related to $\eta_{VE}$, $\eta_{RE}$, $\eta_{IE}$, and $\eta_{LEE}$, respectively. Note that $P_{Joule}$, $P_{NR}$, and $P_{abs}$ have positive values. However, $P_{HC}$ can have a positive or negative value depending on the operating conditions.[17,18] We define $P_{HC}$ as positive (carrier cooling) when carriers release energy to the lattice, and the opposite case as negative (carrier heating).

The total power loss ($P_{loss}$) can be expressed as $P_{loss} = P_{Joule} + P_{NR} + P_{HC} + P_{abs}$. $P_{Joule}$ and $P_{abs}$ can be expressed by

$$P_{Joule} = I^2 R_S ; \qquad (2)$$

$$P_{abs} = \left(\eta_{LEE}^{-1} - 1\right) P_{opt} , \qquad (3)$$

where $R_S$ is the series resistance. However, it is not easy to select an appropriate value of $R_S$ that results from the pure ohmic properties arising at the metal-semiconductor interface, bulk epitaxial layers since space charges are accumulated around the active region during transport.[19, 20] Here, $P_{Joule}$ is calculated with the assumption of emitted photon energy corresponded to junction voltage, i.e., $V_J = (h\lambda_p/c)/q$ and additional voltage equal to difference between junction voltage and external applied voltage, i.e., $\Delta V = V - V_J$, where $h$, $\lambda_p$, $c$ and $q$ are the Planck constant, the peak wavelength, the speed of light, and the elementary charge, respectively. Fig. 2 (a) depicts the *I-V* curves of a sample on linear and semi-log scales. Fig. 2 (b) represents the additional voltage $\Delta V$ as a function of driving current *I*. By using the result in Fig. 2 (b), $P_{Joule}$ can be obtained by $I \cdot \Delta V$. The $\eta_{LEE}$ is estimated as 62% by using the experimental values of the external quantum efficiency ($\eta_{EQE}$) and the internal quantum efficiency ($\eta_{IQE}$). The obtained $\eta_{LEE}$ value is similar to the ones previously reported.[22,23] Using eqs. (1) - (3), we can estimate the three components, $P_{loss}$, $P_{Joule}$, and $P_{abs}$ experimentally.

The total driving current *I* consists of two carrier recombination currents, namely, the radiative recombination current ($I_R$) and the nonradiative carrier recombination current ($I_{NR}$), i.e., $I = I_R + I_{NR}$. The ratio of $I_R$ to *I* is defined as $\eta_{IQE}$, i.e., $\eta_{IQE} = I_R/(I_R + I_{NR})$. The absolute light output power $P_{opt}$ is linearly proportional to $I_R$. The corresponding electrical input power is given as $I_R V_J$. $I_R$, $I_{NR}$, and $V_J$ are found by using the following relations: $I_R = \eta_{IQE} I$, $I_{NR} = (1-\eta_{IQE})I$, and $V_J = V - \Delta V$. Since $P_{NR}$ is experimentally obtained by $I_{NR} V_J$, $P_{HC}$ can be obtained from $P_{loss} - P_{Joule} - P_{abs} - P_{NR}$.[17, 18]

Next, we speculate on the carrier energy loss mechanisms as a function of total driving current by considering In Fig. 3 (a), $P_{Joule}$, $P_{abs}$, $P_{NR}$, and $P_{HC}$ are plotted as a function of total driving current *I*. In Fig. 3 (b), a magnified $P_{HC}$ is shown in low power region in semi-log scale. It should be noted that the sign of $P_{HC}$ is negative in the specific current region. This means that the carries injected into the QWs are energized by the lattice and the lattice is subsequently cooled rather than heated.

For more detailed analysis, we interpret the curve by partitioning it into three regions as denoted in Fig. 3(b). In initial current injection region (R1), $P_{HC}$ is almost zero and constant. In middle current injection region (R2), $P_{HC}$ rapidly decreases from zero to negative values with increasing total current *I*. In high current injection region (R3), $P_{HC}$ continuously increases from the minimum value to positive value with increasing total current *I*.

We interpret experimental curve of Fig. 3 (b) by considering different physical processes of carrier energy losses in three regions as schematically illustrated in Fig. 4. In R1, both the total current level and the carrier density in the MQW region are very small. Both electrons and holes are diffused into the MQW region and most of them are captured in the QWs by transitioning from the 3-dimensional bulk to the 2-dimensional QW electronic states. Then carriers in the QWs lose their energy via radiative or nonradiative recombination processes. Since the carrier is diffused into MQWs by external applied bias and carrier density is determined by their quasi-Fermi level, carrier heating and cooling is unnecessary during this carrier transport. This transport and recombination of carriers is represented by the red dashed line in Fig. 4. As a result, we expect that $|P_{NR}| \approx 0$ and $\eta_{IE}$ of ~100%.

In R2, we think that the carrier density monotonically increases as the carriers accumulate inside the QWs with increasing total driving current, which is due partly to the saturation of the SRH nonradiative recombination rate and partly to the slow increase of the radiative recombination rate compared to the carrier injection rate into the QWs. The saturation of the SRH nonradiative recombination rate results from the limited defect density in the QWs. On the other hand, saturation of the radiative recombination rate induced by phase space filling in the QWs is also expected when the electronic DOS in the conduction and valence bands satisfying the Fermi golden rule is not sufficient.[5, 9, 15] This is considered significant in InGaN-based QWs where the electric potential is localized and randomly fluctuating in space.[4, 5, 24, 25] As shown in Fig. 3 (b), in R2, $P_{HC}$ drastically decreases from zero to negative values. This behavior with increasing *I* can be explained as follows: the SRH nonradiative recombination process becomes saturated because of good crystal quality with a limited defect density in the QWs. As a result, $P_{HC}$ decreases from zero to negative, which means that the total carrier energy increases, indicating the accumulation of carriers at higher energy levels in the QWs. The accumulation of carriers in the QWs originates principally from the fact that the total recombination rate, i.e., the sum of the radiative and nonradiative recombination rates, is smaller than the carrier injection rate. In R2, the nonradiative recombination rate via the SRH recombination process is limited as mentioned above, leading to the radiative recombination rate not meeting the carrier injection rate. This behavior of carrier accumulation is schematically demonstrated as the blue dashed line in Fig. 4.

In R3, $P_{HC}$ increases again from minimum value to positive value with increasing total current *I*. It is expected that $P_{HC}$ has a positive value after the onset of carrier spill-over from the QWs to the p-type AlGaN layer. $P_{NR}$ is mainly due carrier cooling process due to carrier lost their energy to lattice by the carriers overflown from the QWs without being captured there. This behavior of carrier spill-over is schematically demonstrated as green dash line in Fig. 4. Although the SRH recombination is dominant in R1 and R3, the SRH recombination processes occur at different regions, i.e., inside the QWs for R1 ($\eta_{IE}$ of ~100%) and at the p-type clad layers for R3 ($\eta_{IE}$ < 100%).

To investigate the effect of carrier accumulation in MQWs on electrical and optical performance of device, we analyze the ideality factor and IQE data by partitioning it into three regions as same in Fig. 3(b). In Fig. 5 (a), ideality factor is plotted as function of total driving current. The ideality factor (*n*) is expressed as,[26]

$$n = \frac{q}{kT}\left(\frac{d\ln I}{dV}\right)^{-1}, \qquad (4)$$

where *q* is elementary charge, *k* is Boltzmann constant, and *T* is absolute temperature. It is typical ideality factor characteristic as a function of driving current in InGaN-based LED. In particular, the current at minimum value of *n* ($I_{min}$) indicate the onset of high-level-injection, namely, low-level-injection is dominant at $I_{min} > I$ and high-level-injection is dominant at $I_{min} < I$.[10] As shown in Fig. 5 (a), the onset of high-level-injection is fairly similar to the transition from R1 to R2. It means that carrier accumulation lead to condition of high-level injection such as $n_p \approx p_p$ around active region, which induce the charge building up and additional potential drop.[19] It is also thought to be one of possible origin of efficiency droop,[10] thus, we can infer that this charge accumulation significantly effect on LED device performance.[20] In Fig. 5 (b), $\eta_{IQE}$ is plotted as function of current. In R1, $\eta_{IQE}$ increases monotonically up to ~60% and in R2, $\eta_{IQE}$ increases to its peak value. In R3, $\eta_{IQE}$ monotonically decreases with *I,* which is known as the "efficiency droop" phenomenon. Based-on our model, experimental result of $\eta_{IQE}$ can be explained as followings: at R1, $\eta_{IQE}$ is less than 60% and most of the carriers are diffused and captured into the QWs. In this current range, carriers recombine sufficiently via both the radiative and the nonradiative recombination ($\eta_{IE} \approx 100\%$). When $\eta_{IQE}$ is between ~55% and the peak value in the experimental sample (R2), both recombination rates do not increase sufficiently compared to the carrier injection rate, resulting in carriers being accumulated in the QWs and eventually spilt-over from the QWs to the p-type AlGaN layer. At the high injection current level (R3), carriers are continuously overflown from the QWs to the p-type clad layer and lose their electrical energy thermally and it induce efficiency droop. From above analysis, we can see that carrier accumulation process begins from a relatively low current density.

In conclusion, we investigate the dominant carrier transport and recombination processes inside and outside the MQW region as a function of injection current from a new point of view by separately examining the carrier energy loss. Based on our model, the radiative and nonradiative carrier recombination processes can be distinctively understood in terms of three current injection regions. Analysis of the measurement results reveals that the carrier accumulation and subsequent spill-over from the MQW active region to the clad is the most probable mechanism of explaining the efficiency and forward voltage variation with the injection current.

**Figures captions**

FIG. 1. $P_{elec}$, $P_{opt}$, $P_{loss}$, and $\eta_{WPE}$ characteristics measured as a function of total driving current.

FIG. 2. (a) *I-V* characteristics of sample on linear and semi-log scale and (b) $\Delta V$ plotted as a function of driving current.

FIG. 3. (a) $P_{loss}$, $P_{Joule}$, $P_{abs}$, $P_{NR}$, and $P_{HC}$ measured as a function of total driving current and (b) $P_{HC}$ magnified at the low power region in semi-log scale.

FIG. 4. Schematic illustration of the band diagram including the proposed carrier transport and recombination mechanisms in the LED device.

FIG. 5. (a) Ideality factor and (b) IQE plotted as a function of driving current in semi-log scale.

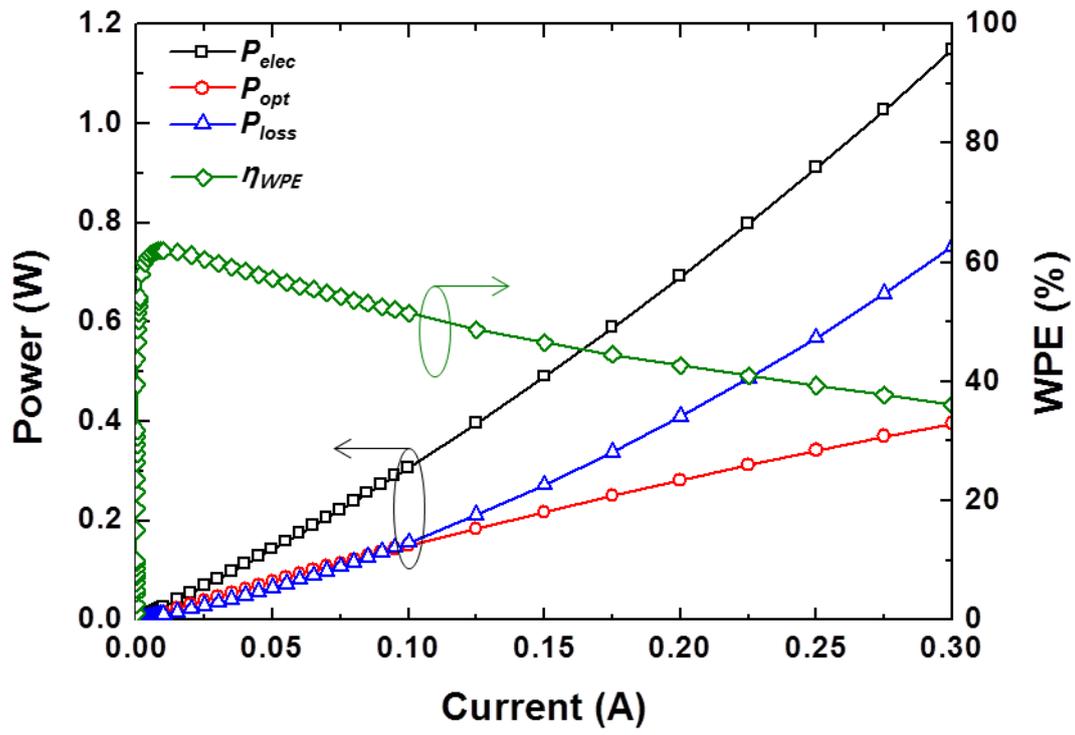

Fig. 1

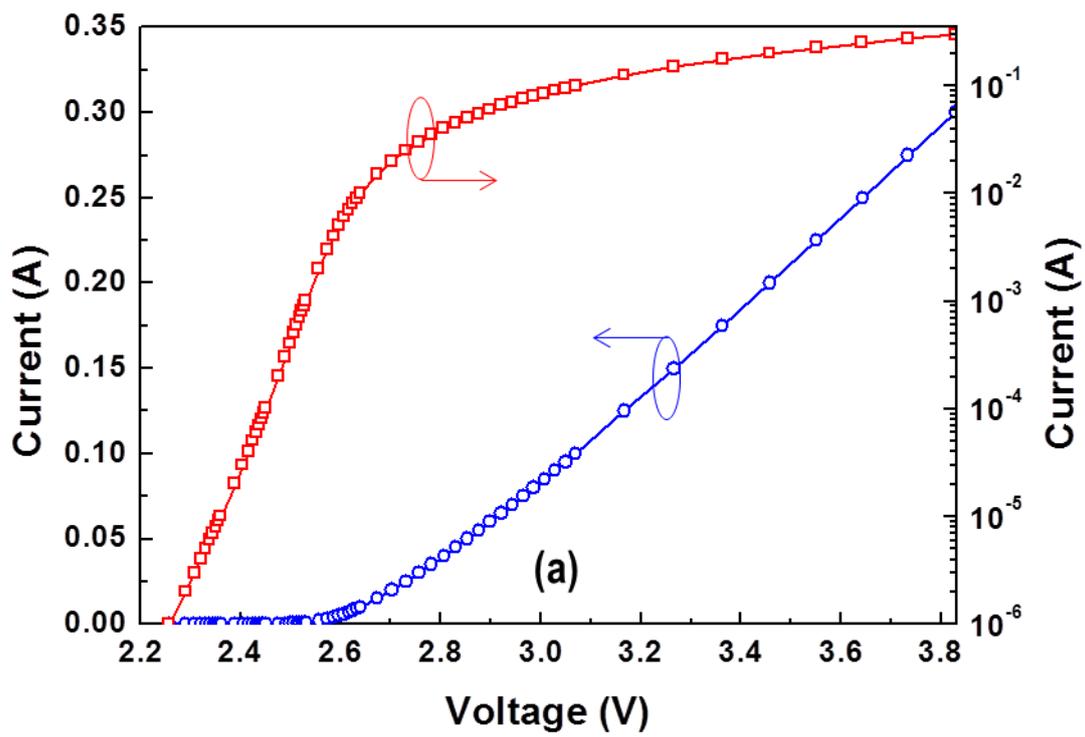

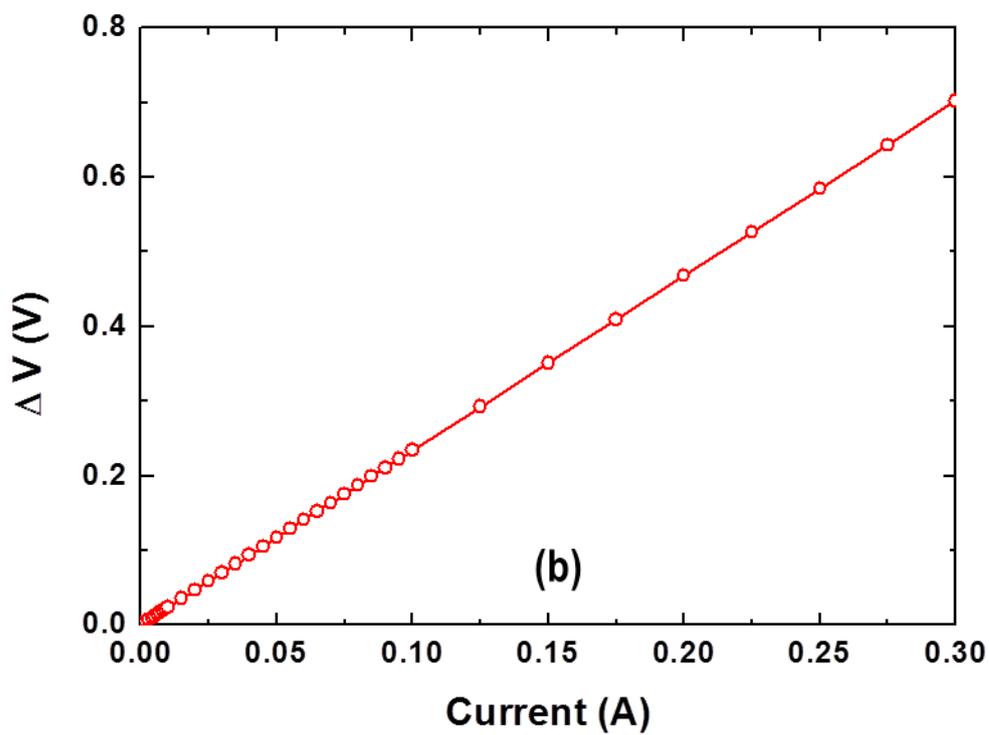

Fig. 2

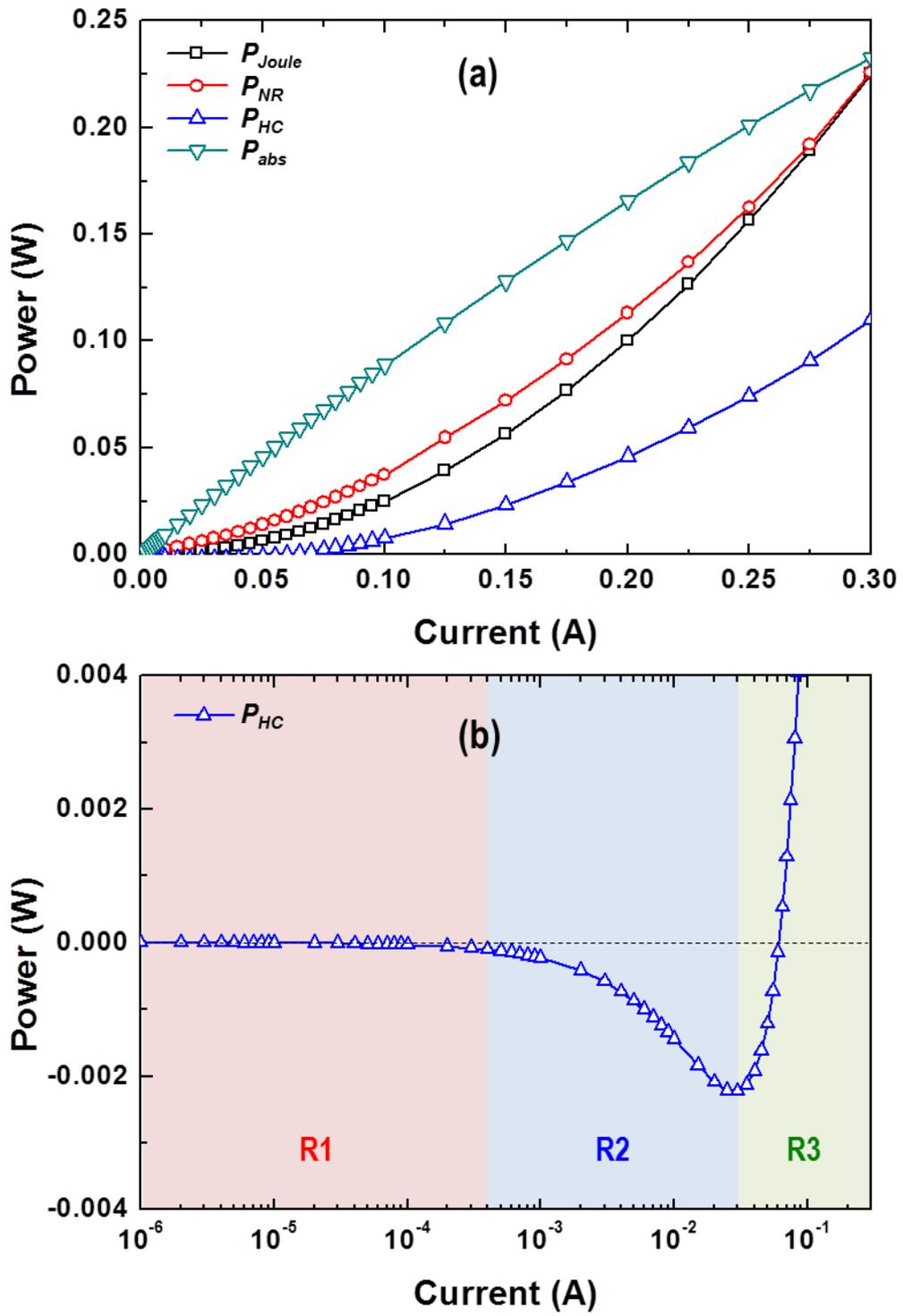

Fig. 3

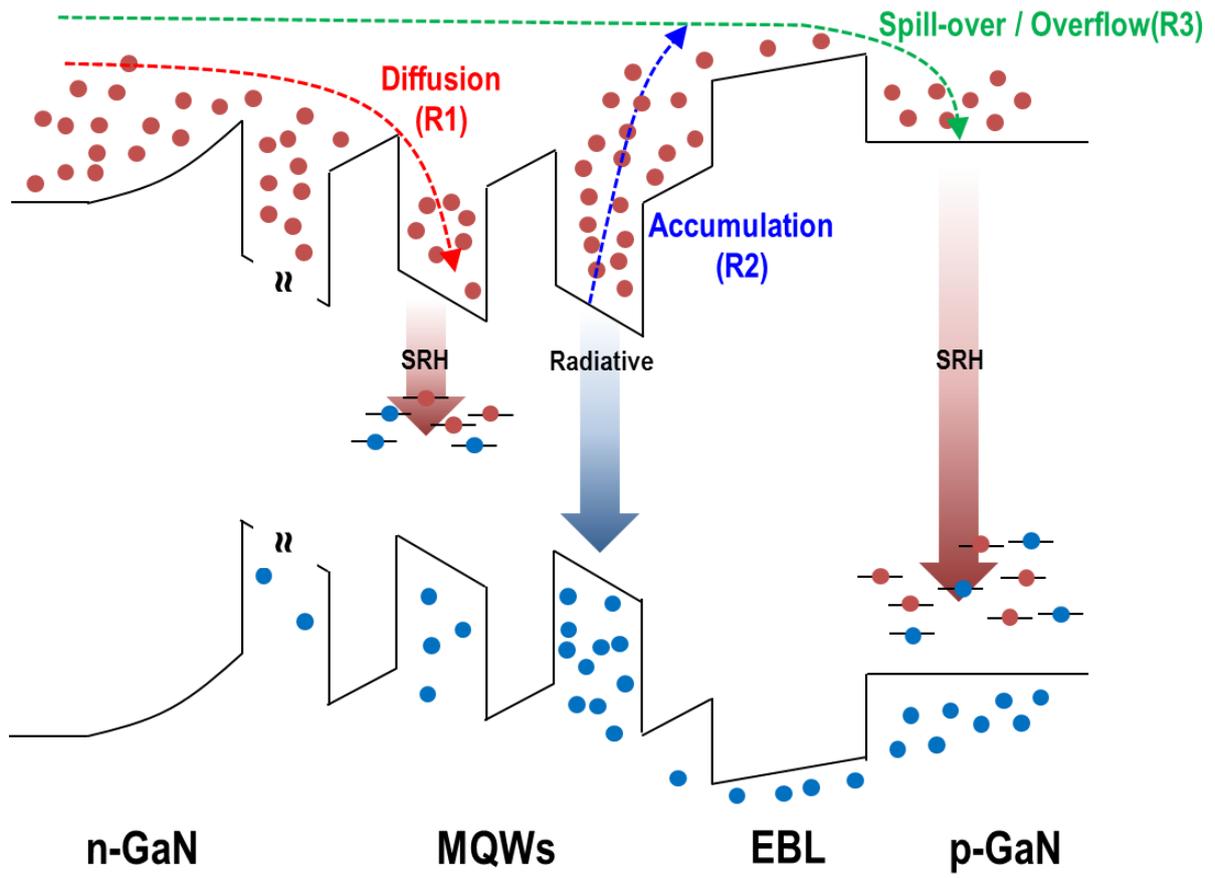

Fig. 4

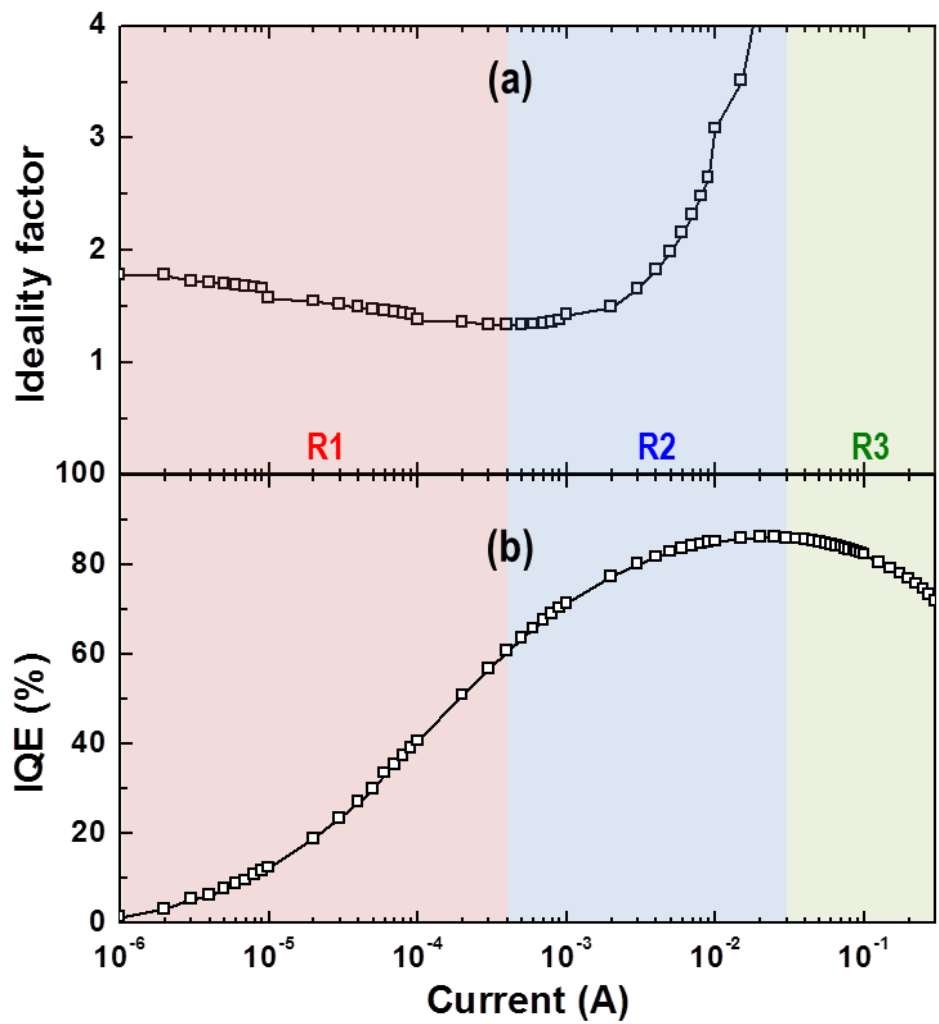

Fig.5